\documentclass[letterpaper]{article}
\usepackage{aaai22}  
\usepackage{times}  
\usepackage{helvet} 
\usepackage{courier}  
\usepackage[hyphens]{url}  
\usepackage{graphicx} 
\usepackage{booktabs}
\usepackage{natbib}  
\usepackage{caption} 
\usepackage{cleveref}
\usepackage{multicol}
\usepackage{subcaption}
\usepackage{todonotes}

\begin{document}

\title{A Psycho-linguistic Analysis of BitChute \\
[1ex] \large A Metadata Supplement to The MeLa BitChute Dataset \\}
\author {
    Benjamin D. Horne \\
}
\affiliations {
  School of Information Sciences, University of Tennessee Knoxville, Knoxville, TN, USA\\
  bhorne6@utk.edu\\
}

\maketitle

\begin{abstract}
In order to better support researchers, journalist, and practitioners in their use of the \texttt{MeLa-BitChute} dataset for exploration and investigative reporting, we provide new psycho-linguistic metadata for the videos, comments, and channels in the dataset using LIWC22. This paper describes that metadata and methods to filter the data using the metadata. In addition, we provide basic analysis and comparison of the language on BitChute to other social media platforms. The \texttt{MeLa-BitChute} dataset and LIWC metadata described in this paper can be found at: \url{https://dataverse.harvard.edu/dataset.xhtml?persistentId=doi:10.7910/DVN/KRD1VS}.
\end{abstract}


\section{Introduction}
The alt-tech ecosystem, a set social media platforms that exist in answer to perceived risks of censorship from large social media platforms, has created a digital infrastructure for fringe groups, particularly on the far-right \cite{jasser2021welcome, donovan2019parallel,Wilson2021}. Platforms in this ecosystem have provided many technological affordances to these groups, such as low content moderation, mechanisms to grow engaged audiences, and sometimes even funding structures for content production \cite{jasser2021welcome, trujillo2020bitchute}. 

Due to these affordances, a primary concern with the continued growth of alt-tech platforms is, for lack of a better term, the \textit{offline harms} that are facilitated or incited by online activities and extremist movements on those platforms \cite{munn2021more}. For example, it has been argued that violent events such as the 2016 Comet Ping Pong pizzeria gunman (\textit{Pizzagate}), the 2017 Unite the Right rally in Charlottesville, and the 2021 U.S. Capitol attack have each had online components, ranging from organization, coordination, and inspiration. Although, the effect of social media on ideology, events, and actions is widely debated \cite{guess2018avoiding, flaxman2016filter, althoff2017online, rice2022monitoring}, gaining a better understanding of what types of calls to violence exist online and the dynamics involved in potentially violent movements is still salient. 

Work by qualitative, ethnographic researchers and investigative journalists is critical in gaining this understanding. Often, in quantitative, big data research, we focus on the the \textit{elite}, highly-productive, and highly-engaged with content producers in a space, sometimes missing the smaller players, who may still generate consequential harms both online and offline. Yet, filtering large datasets to smaller datasets suitable for qualitative work is time-consuming and can be a barrier-to-entry to studying niche, yet consequential, behaviors on large social platforms. 

Released in early 2022, the \texttt{MeLa-BitChute} dataset \cite{trujillo2022mela} provides a large, near-complete sample of data from 3M+ videos, 11M+ comments, and 61K+ channels on one such alt-tech platform, BitChute. Given the structure of the dataset, it is suitable for large-scale studies of the platform out-of-the-box, but requires some additional effort to perform small-scale, qualitative studies of the platform. To better facilitate and support qualitative studies and explorations of the platform, we provide a psycho-linguistic metadata set over the \texttt{MeLa-BitChute} dataset using LIWC-22. In this short paper, we describe this metadata set, describe several use cases, and provide practical guidance on using it. 

Both the original \texttt{MeLa-BitChute} dataset and the metadata described in this paper can be found in the following repository: \url{https://dataverse.harvard.edu/dataset.xhtml?persistentId=doi:10.7910/DVN/KRD1VS}. The paper documenting the original dataset collection and structure can be found in \cite{trujillo2022mela}. 

\section{Linguistic Inquiry and Word Count}
Linguistic Inquiry and Word Count (LIWC) is a theory-driven, dictionary-based method to measure various psychological states from open-text, dating back to 1993 \cite{francis1993linguistic}, conceptually stemming from work in Psychology from 1942 \cite{allport1942use}. The method has been improved upon over time, with updates in 2001 \cite{pennebaker2001linguistic}, 2007 \cite{pennebaker2007linguistic}, 2015 \cite{pennebaker2015development}, and 2022 \cite{boyddevelopment}. 

The method has also been used widely across various academic studies and settings. These include studies of social media \cite{eichstaedt2018facebook,coppersmith2014measuring,schwartz2013characterizing}, news media \cite{horne2017just,shu2019beyond}, online reviews \cite{del2014study}, spam detection \cite{crawford2015survey}, conversations \cite{cannava2018stuff}, conference calls \cite{larcker2012detecting}, college admissions essays \cite{pennebaker2014small}, and more. 

The high-level idea is that given a set of normalized, stemmed words grouped into meaningful categories, such as negative emotion, conflict, or affiliation, one can count the occurrence of those words in a document to quickly assess what is being discussed in a document and how. In this work, we propose using this method as a mechanism for filtering and exploring the \texttt{MeLa-BitChute} dataset. By computing each LIWC category across all video titles, comments, and aggregating those scores by channels, we can effectively search for various types of content in the dataset, rather than searching using single keywords or manually exploring content across the many channels and videos.  

To construct LIWC metadata, we use the latest version of LIWC: LIWC-22. Further documentation on LIWC22 can be found at \url{https://www.liwc.app/}, including definitions of all 117 categories. Below, we describe some examples using these categories, but do not define all categories included in the metadata. 

\section{Metadata Structure}
Just as with the \texttt{MeLa-BitChute} dataset, we provide two widely-used data formats. 

\subsection{SQLite3 Database}

\begin{figure*}[ht!]
    \centering
    \includegraphics[width=17.7cm]{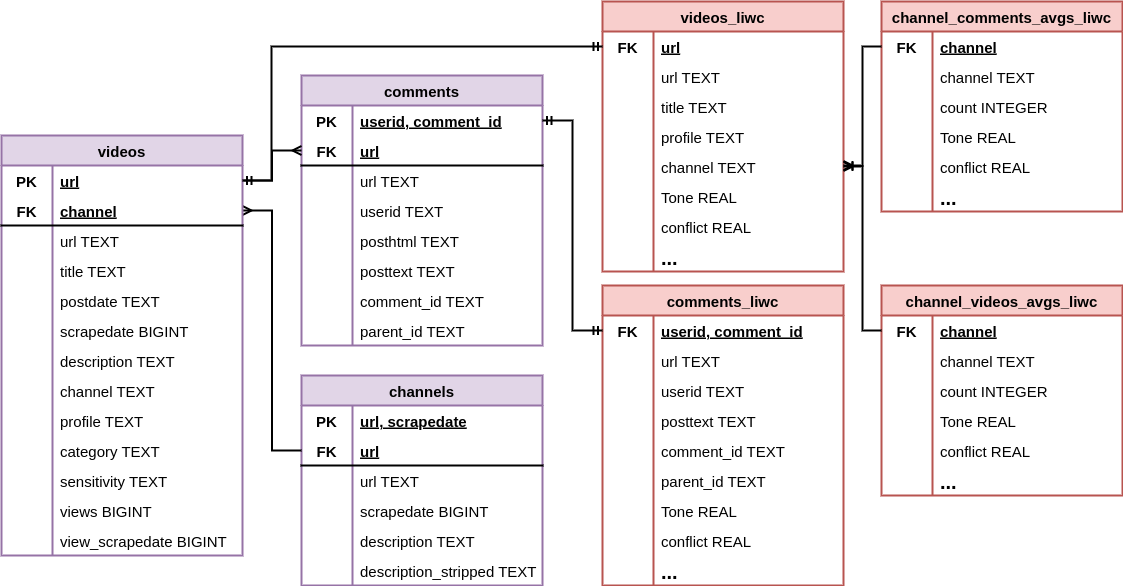}
    \caption{Metadata schema and original data schema. The original dataset tables are in purple, while the new metadata set tables are in red. The \texttt{MeLa-BitChute} dataset and the metadata set are stored in standalone databases. To this end, the metadata tables include the text (titles or comment text) to allow for its use without the original dataset. Note, each metadata table contains 115 more LIWC category columns that are not shown to save space.}
    \label{fig:schema}
\end{figure*}

The first format is an SQLite3 database with four tables: 
\begin{enumerate}
    \item \texttt{videos\_liwc} - This table contains the video URL, title, profile, and channel, along with 117 LIWC categories calculated on each video title. 
    \item \texttt{comments\_liwc} - This table contains anonymized user ID, video URL, comment ID, and parent ID, along with 117 LIWC categories calculated on each comment. User IDs are salted hashes of each user's account information, allowing for comments to be grouped by users  without revealing the username of the author. For more details on comment completeness and ID creation, see \cite{trujillo2022mela}. 
    \item \texttt{channel\_comments\_avgs\_liwc} - This table contains the URL to the channel, the number of comments made on videos by the channel (called count), and the average of each LIWC category across all comments made on videos by the channel. 
    \item \texttt{channel\_videos\_avgs\_liwc} - This table contains the URL to the channel, the number of videos by the channel (called count), and the average of each LIWC category across all video titles by the channel. 
\end{enumerate}

In Figure \ref{fig:schema}, we show both the \texttt{MeLa-BitChute} dataset schema and the LIWC metadata schema.

While both the original dataset and the metadata set can be used together, we choose to store each in independent databases for ease of use. Specifically, the video URLs, comment IDs and channel URLs can all be mapped back to the \texttt{MeLa-BitChute} dataset, but the needed text and URLs are also stored in the metdata set, allowing for exploration without joining the two databases. In Figure \ref{fig:schema}, we show how the two sets are related.

\subsection{CSV}\label{sec:csv}
The second format in which we provide the dataset is a set of Comma-Separated Value (CSV) files. We provide four CSV files, one for each table in the database. The columns in each CSV file are the same as the columns in each corresponding SQLite3 database table. 

\section{Use Cases}
There are several ways the \texttt{MeLa-BitChute} dataset can be explored using this metadata. 

\subsection{BitChute Compared to Itself and Other Social Platforms}
First, using LIWC22 categories we can quickly examine if a channel or set of channels are producing content that is like other social platforms or not. In Table \ref{tab:compare}, we show the average and standard deviation of 17 LIWC categories across all of BitChute. These averages can be used as baselines to compare BitChute to other social platforms. For example, using Table \ref{tab:compare}, we can see that on average BitChute comments use more `negative tone' (words such as bad, wrong, hate, etc.) than other social platforms such as Facebook, Reddit, Twitter, and online Blogs. We also see that the dispersion of negative tone scores across comments on BitChute is much higher than other social platforms. Similarly, we can see a higher use of ethnicity (words like Jew, American, French, Chinese, Indian, etc.) and religion (words like god, hell, christmas, church, etc.) in the comments on average than other platforms. When looking at the video titles, we see a higher use of conflict (words like fight, kill, killed, attack, etc.) and political words (words like United States, govern, congress, senate, etc.) on average than other platforms. 

Second, we can use these LIWC baselines to compare individual channels to the rest of BitChute. For example, using the \texttt{channel\_videos\_avgs\_liwc} database table, we see that the channel `banned-dot-video', one of several Infowars channels on BitChute, uses more conflict words (1.79), power words (6.13), death words (1.16), and negative tone (5.43) in video titles than the rest of BitChute on average. Similarly, we can see video titles for channels like `zionistreport', one of several anti-Semitic channels on BitChute, use more conflict words (0.97), ethnicity words (6.96), power words (7.04), death words (1.10), affiliation words (2.74), emotional anger (0.64) and negative tone (4.31) on average than the rest of BitChute video titles.

\begin{table*}[ht!]
    \centering
    \fontsize{10pt}{11.8pt}
    \selectfont
    \begin{tabular}{c|c|c|c|c|c|c}
        \toprule
        \textbf{LIWC22 Category} & \textbf{BitChute Videos} & \textbf{BitChute Comments} & \textbf{Facebook} & \textbf{Reddit}& \textbf{Twitter} & \textbf{Blogs}\\\midrule
        \textbf{Affect} & 5.69 $\pm$ 10.44 & \textcolor{red}{\textbf{9.79 $\pm$ 13.71}} & 8.82 $\pm$ 2.47 & 5.72 $\pm$ 1.70 & 8.96 $\pm$ 4.48 & 5.54 $\pm$ 1.64\\
        \textbf{emo\_anger} & 0.16 $\pm$ 1.84 & 0.18 $\pm$ 1.59 & \textcolor{red}{\textbf{0.22 $\pm$ 0.23}} & 0.19 $\pm$ 0.25 & 0.18 $\pm$ 0.20 & 0.18 $\pm$ 0.26\\
        \textbf{emo\_neg} & 0.58 $\pm$ 3.42 & 1.24 $\pm$ 4.03 & \textcolor{red}{\textbf{1.29 $\pm$ 0.89}} & 0.79 $\pm$ 0.53 & 0.76 $\pm$ 0.56 & 0.81 $\pm$ 0.59\\
        \textbf{ton\_neg} & 2.93 $\pm$ 7.34 & \textcolor{red}{\textbf{3.99 $\pm$ 8.00}} & 2.34 $\pm$ 1.14 & 2.10 $\pm$ 0.94 & 1.85 $\pm$ 1.07 & 1.76 $\pm$ 0.93\\
        \textbf{swear} & 0.29 $\pm$ 2.64 & \textcolor{red}{\textbf{2.40 $\pm$ 7.99}} & 0.52 $\pm$ 0.72 & 0.71 $\pm$ 0.67 & 1.08 $\pm$ 1.42 & 0.33 $\pm$ 0.53\\
        \textbf{prosocial} & 0.31 $\pm$ 2.43 & 0.74 $\pm$ 4.02 & 0.69 $\pm$ 0.58 & 0.47 $\pm$ 0.45 & \textcolor{red}{\textbf{1.17 $\pm$ 0.97}} & 0.44 $\pm$ 0.35\\
        \textbf{conflict} & \textcolor{red}{\textbf{0.82 $\pm$ 3.80}} & 0.65 $\pm$ 3.13 & 0.22 $\pm$ 0.23 & 0.35 $\pm$ 0.35 & 0.27 $\pm$ 0.25 & 0.23 $\pm$ 0.26\\
        \textbf{politic} & \textcolor{red}{\textbf{1.35 $\pm$ 5.17}} & 1.15 $\pm$ 3.88 & 0.11 $\pm$ 0.26 & 0.26 $\pm$ 0.45 & 0.42 $\pm$ 0.92 & 0.29 $\pm$ 0.72\\
        \textbf{ethnicity} & 0.77 $\pm$ 3.98 & \textcolor{red}{\textbf{0.90 $\pm$ 3.89}} & 0.11 $\pm$ 0.19 & 0.18 $\pm$ 0.32 & 0.16 $\pm$ 0.32 & 0.13 $\pm$ 0.28\\
        \textbf{female} & 0.42 $\pm$ 2.90 & 0.61 $\pm$ 3.06 & 0.72 $\pm$ 0.70 & \textcolor{red}{\textbf{0.93 $\pm$ 1.01}} & 0.79 $\pm$ 0.63 & 0.92 $\pm$ 1.10\\
        \textbf{relig} & 1.19 $\pm$ 5.15 & \textcolor{red}{\textbf{1.36 $\pm$ 5.49}} & 0.54 $\pm$ 0.68 & 0.27 $\pm$ 0.39 & 0.53 $\pm$ 1.16 & 0.34 $\pm$ 0.65\\
        \textbf{moral} & 0.70 $\pm$ 3.70 & \textcolor{red}{\textbf{1.04 $\pm$ 4.38}} & 0.27 $\pm$ 0.25 & 0.36 $\pm$ 0.36 & 0.40 $\pm$ 0.40 & 0.28 $\pm$ 0.26\\
        \textbf{death} & \textcolor{red}{\textbf{0.73 $\pm$ 3.61}} & 0.53 $\pm$ 2.70  & 0.18 $\pm$ 0.22 & 0.23 $\pm$ 0.28 & 0.17 $\pm$ 0.26 & 0.11 $\pm$ 0.17\\
        \textbf{sexual} & 0.21 $\pm$ 2.12 & \textcolor{red}{\textbf{0.41 $\pm$ 3.09}} & 0.09 $\pm$ 0.19 & 0.29 $\pm$ 0.46 & 0.13 $\pm$ 0.23 & 0.11 $\pm$ 0.29\\
        \textbf{affiliation} & 1.24 $\pm$ 4.70 & 1.47 $\pm$ 4.10 & 1.72 $\pm$ 0.92 & 1.53 $\pm$ 0.90 & \textcolor{red}{\textbf{2.31 $\pm$ 1.47}} & 1.93 $\pm$ 1.12 \\
        \textbf{power} & \textcolor{red}{\textbf{3.01 $\pm$ 7.29}} & 2.36 $\pm$ 5.50 & 0.74 $\pm$ 0.54 & 1.13 $\pm$ 0.77 & 1.22 $\pm$ 1.15 & 0.93 $\pm$ 0.82\\
        \textbf{we} & 0.46 $\pm$ 2.60 & 0.73 $\pm$ 2.59 & 0.61 $\pm$ 0.53 & 0.54 $\pm$ 0.53 & \textcolor{red}{\textbf{0.97 $\pm$ 1.02}} & 0.91 $\pm$ 0.83\\\bottomrule
    \end{tabular}
    \caption{Mean and Standard Deviation ($\mu \pm \sigma$) of selected LIWC22 categories across social platforms. Highlighted in bold red are the highest averages in each row. The column `BitChute Videos' is the average LIWC category score across 3,036,190 video titles and the column `Bitchute Comments' is the average LIWC category score across 11,434,571 comments. The columns for Facebook, Reddit, Tweets, and Blogs are from the LIWC22 Test Kitchen Corpus (See here: \url{https://www.liwc.app/static/documents/LIWC-22.Descriptive.Statistics-Test.Kitchen.xlsx}). A CSV file with all LIWC22 categories can be found on Dataverse.}
    \label{tab:compare}
\end{table*}

\subsection{Ranking Channels, Comments, and Videos}
These LIWC categories can also be used to rank channels by various word usages. For example. in Figures \ref{fig:ethnicity}, \ref{fig:toneneg}, \ref{fig:conflict}, and \ref{fig:power} in Appendix D, we show rankings of channels by their average use of a LIWC category in the comments or video titles. 

Using this ranking method we can find channels that have audiences who use high amounts of ethnicity words in the comments, pointing to channels such as `phoenix\_party\_fascists' - an anti-Semitic channel that has since been blocked by BitChute due to `Platform Misuse'. Platform Misuse is a somewhat recent addition to the BitChute community guidelines - first appearing on the website in mid 2020. It states that channels can be blocked for behaviors such as brigading, metric manipulation, name squatting, scamming, or spamming. Importantly, it appears this channel was not blocked due to its anti-Semitic hate speech, but rather one of the listed platform misuses. 

This ranking method also find channels with particular psychological drives, such as use of power word words like own, order, allow, power, etc.) in the video titles. For example, in Figure \ref{fig:power} the top channel is Steve Bannon's `pandemic war room' - a channel that publishes Steve Bannon's radio shows discussing everything from anti-intellectualism to COVID-19 conspiracy theories. 

Importantly, when ranking by LIWC categories, one should provide a threshold for the number of comments or videos. For instance, if a channel has one video who's title uses all negative tone words, than it will have a `tone\_neg' score of 100 on average. However, since the channel only produced one video, being ranked highly in negative tone is probably not very meaningful. Instead, if we use the `count' column in the database, we can filter to only rank channels that have more than a certain number of videos. See Table \ref{tab:example_SQL} in Appendix B for an example of this filter in SQL. 

\subsection{Exploring Topical Focuses on the platform}
Several of the LIWC categories are topical in nature. For example, the categories `politic' and `relig' can show us what channels discuss politics and what channels discuss religion. When examining the rankings in Figure \ref{fig:relig} and \ref{fig:politics} in Appendix D, we can quickly see the top channels in each topic. For discussion of politics, we see channels such as `OANN', the well-known far-right news network, and `DonaldJTrump', a channel that publishes Donald Trump's speeches. For discussion of religion, we see channels such as `StephenKJV1611', the channel of Stephen Anderson\footnote{Stephan Anderson is known for anti-homosexual hate speech and has been banned from several countries, read more here: \url{https://en.wikipedia.org/wiki/Steven_Anderson_(pastor)}}, and the channel `Church-Militant', a claimed Catholic faith channel containing a variety of conspiracy theories.

\section{Recommended Tools and Methods for Exploring using the Metadata}
Given the large size of the dataset and the complexity of various categories in LIWC, we recommend exploring and filtering the data using SQL. For those unfamiliar with SQL, we provide some plug-n-play examples of SQL statements for the metadata in Table \ref{tab:example_SQL} in Appendix B. 

Furthermore, we recommend using SQLite DB Browser for easy exploration\footnote{\url{https://sqlitebrowser.org/}}. In Figure \ref{fig:dbbrow} in Appendix C, we show screenshots of executing SQL and filtering columns by keywords in DB Browser. 
While we do provide the CSV versions of this metdata for use, it is likely too large to effectively explore in software like Excel on a typical laptop, while a database browser can handle the large size by not loading all the data at once. 

\section{Limitations}
There are several important limitations of dictionary-based methods like LIWC that should be kept in mind when using this metadata. 

First, the dictionaries are language specific. While the vast majority of BitChute is in English \cite{trujillo2020bitchute}, some channels are not. If the channel is not in English, the LIWC category values cannot be relied on. For example, if a channel is in German, the death category in LIWC will be high, as the German word for `the' is `die'. This limitation may slightly inflate the average use of `death' words across the platform. 

Second, it is well known that in fringe communities, language may be used in community-specific ways not captured by LIWC. While LIWC has an extensive `netspeak' category, it is unlikely this covers all of the \textit{dog-whistles} and coded language used by fringe groups. For example, the use of triple parenthesis around a word, such as (((jew))) or (((they))), in fringe communities often refers to anti-Semitic conspiracy theories and contexts \cite{zannettou2020quantitative}. These types of coded languages are not captured by LIWC. Although the word `jew' appears to be captured in both the ethnicity and religion categories, words such as `they' are simply captured as 3rd person plural words.

Third, while LIWC has been validated in many settings, dictionary-based methods naturally lose the context around individual words. For example, two comments may use the same word in the category `death' - one comment may be a call to violence, while the other may be discussing the Biblical theology of death. These contextual differences should be taken into account when interpreting aggregate results. This limitation has also been noted in other studies \cite{hirsh2009personality,bantum2009evaluating}.

\section{Conclusion}
In this paper, we describe a LIWC metdata set for use in exploration and sub-setting the \texttt{MeLa-BitChute} dataset. We provide multiple levels of metadata, including LIWC scores for video titles, comments, and aggregations of both per channel. In addition, we provide averages of each category across the full platform to provide baselines for comparing channels to the rest of BitChute and to other social media platforms. Lastly, we provide example plug-in-play SQL statements for exploring the metadata and a guide to using the SQLite DB Browser. 

Our hope is that this metadata can better support qualitative researchers and investigative journalist in the use of the \texttt{MeLa-BitChute} dataset, and that it can provide LIWC baselines for researchers to compare other alt-tech platforms to BitChute.

Both the original \texttt{MeLa-BitChute} dataset and the metadata described in this paper can be found in the following repository: \url{https://dataverse.harvard.edu/dataset.xhtml?persistentId=doi:10.7910/DVN/KRD1VS}

\appendix
\onecolumn
\section{Data Column Descriptions}\label{appendix:coumns}
In this section, we provide descriptions of each data column in the \texttt{MeLa-BitChute} supplemental metadata. Below are tables for each table in the database (videos\_liwc, comments\_liwc, channels\_comments\_avg\_liwc, and channels\_videos\_avg\_liwc). Note, to save space, we do not list each LIWC category. However, all 117 LIWC categories are included with the same names as provided by LIWC. For a detailed description of each, please see \url{https://www.liwc.app/}.

\begin{table*}[ht!]
\fontsize{9pt}{9pt}
\selectfont
\centering
\begin{tabular}{c|p{11cm}}
\textbf{Column Name} & \textbf{Description}\\
\toprule
url & URL to video\\\midrule
title & Title of the video\\\midrule
profile & URL to the uploader's profile. Note, a profile can have multiple channels, but a channel belongs to one profile.\\\midrule
channel & URL to the channel \\\midrule
All LIWC categories & 117 columns corresponding to each LIWC category. The column names are the same as the names found in the LIWC documentation. Each LIWC category is a number between 0 and 100, representing a percent of text that falls in that category. Please note baselines can vary widely for LIWC categories based on the size of the dictionary.\\\bottomrule
\end{tabular}
\caption{\textbf{videos\_liwc} data description.}
\label{tbl:vids}
\end{table*}

\begin{table*}[ht!]
\fontsize{9pt}{9pt}
\selectfont
\centering
\begin{tabular}{c|p{11cm}}
\textbf{Column Name} & \textbf{Description}\\
\toprule
url & URL to video that the comment falls under\\\midrule
userid & A SHA256 hash that uniquely identifies each commenter \\\midrule
posttext & The body text of the comment (a pre-processed version of \texttt{posthtml} in the original dataset) \\\midrule
comment\_id & A text ID identifying a comment on a video \\\midrule
parent\_id & If non-NULL, refers to the \texttt{comment\_id} of the parent comment\\\midrule
All LIWC categories & 117 columns corresponding to each LIWC category. The column names are the same as the names found in the LIWC documentation. Each LIWC category is a number between 0 and 100, representing a percent of text that falls in that category. Please note baselines can vary widely for LIWC categories based on the size of the dictionary.\\\bottomrule
\end{tabular}
\caption{\textbf{comments\_liwc} data description}
\label{tbl:cmts}
\end{table*}

\begin{table*}[ht!]
\fontsize{9pt}{9pt}
\selectfont
\centering
\begin{tabular}{c|p{11cm}}
\textbf{Column Name} & \textbf{Description}\\
\toprule
channel & URL to the channel\\\midrule
count & Number of comments on videos by the channel \\\midrule
Average of All LIWC categories & The average LIWC score of comments on videos by the channel, done for all 117 LIWC categories. The column names are the same as the names found in the LIWC documentation.\\\bottomrule
\end{tabular}
\caption{\textbf{channels\_comments\_avg\_liwc} data description}
\label{tbl:cmts_db}
\end{table*}

\begin{table*}[ht!]
\fontsize{9pt}{9pt}
\selectfont
\centering
\begin{tabular}{c|p{11cm}}
\textbf{Column Name} & \textbf{Description}\\
\toprule
channel & URL to the channel\\\midrule
count & Number of videos by the channel \\\midrule
Average of All LIWC categories & The average LIWC score of video titles by the channel, done for all 117 LIWC categories. The column names are the same as the names found in the LIWC documentation. \\\bottomrule
\end{tabular}
\caption{\textbf{channels\_videos\_avg\_liwc} data description}
\label{tbl:cmts_db}
\end{table*}

\pagebreak
\section{SQL Examples}\label{appendix:sql}
In the section, we provide several example SQL statements that can be used to explore the dataset. In each example, LIWC categories can be replaced by any other LIWC category. 

\begin{table*}[ht!]
    \centering
    \fontsize{12pt}{11.8pt}
    \begin{tabular}{p{8.4cm}|p{8.4cm}}
    \toprule
       \textbf{SQL statement}  & \textbf{Description} \\\midrule
       SELECT channel, title, ethnicity FROM videos\_liwc WHERE WC $>=$ 5 ORDER BY ethnicity DESC LIMIT 500  & Returns the channel url, video title, and ethnicity LIWC score ranked by the highest use of ethnicity words in the title, where title has at least 5 words. The LIWC category `ethnicity' can be replaced with any LIWC category. We recommend limiting your output when exploring due to the large size of what will be returned. \\\\\midrule
      SELECT url, posttext, ethnicity FROM comments\_liwc WHERE WC $>=$ 10 ORDER BY ethnicity DESC LIMIT 500  & Returns the video url, comment text, and ethnicity LIWC score ranked by the highest use of ethnicity words in the comment, where comment has at least 10 words. The LIWC category `ethnicity' can be replaced with any LIWC category.\\\\\midrule
      SELECT channel, power FROM channel\_video\_avgs\_liwc WHERE count $>=$ 1000 ORDER BY power DESC LIMIT 500 & Returns channels ranked by average use of power words in video titles where the channel has at least 1000 videos. The LIWC category `power' can be replaced with any LIWC category.\\\\\midrule
      SELECT channel, conflict, ethnicity, tone\_neg, power, death, emo\_anger from channel\_video\_avgs\_liwc WHERE channel = '/channel/zionistreport/' & Returns average LIWC scores for conflict, ethnicity, negative tone, power, death, and emotional anger from video titles produced by the channel `zionistreport'. LIWC categories and channel name can be replaced with desired categories and channel name.\\\\\midrule
      SELECT channel, conflict, ethnicity, tone\_neg, power, death, emo\_anger from channel\_comments\_avgs\_liwc WHERE channel = '/channel/banned-dot-video/' & Returns average LIWC scores for conflict, ethnicity, negative tone, power, death, and emotional anger from comments under videos by the channel `banned-dot-video'. LIWC categories and channel name can be replaced with desired categories and channel name.\\\\\midrule
      SELECT channel, conflict FROM channel\_comments\_avgs\_liwc WHERE count $>=$ 100 ORDER BY conflict DESC LIMIT 500 &  Returns channels ranked by average use of conflict words in comments where the channel has at least 100 comments. The LIWC category `conflict' can be replaced with any LIWC category. \\\\\midrule
      SELECT videos\_liwc.channel, comments\_liwc.posttext, comments\_liwc.death FROM comments\_liwc JOIN videos\_liwc ON comments\_liwc.url $=$ videos\_liwc.url WHERE comments\_liwc.WC $>=$ 100 ORDER BY comments\_liwc.death DESC LIMIT 500 & Returns the channel url and comment text ranked by the number of death words in a single comment, where the comment contains at least 100 words. The LIWC category `death' can be replaced with any LIWC category.\\\\\midrule
      SELECT url, posttext, ethnicity from comments\_liwc WHERE ethnicity $>$ 0.90 & Returns all comments that contain more ethinicity words than the average BitChute comment.\\\\\midrule
      SELECT channel, url, conflict from videos\_liwc WHERE conflict $>$ 0.82 & Returns all videos with titles that contain more conflict words than the average BitChute video.\\\\\bottomrule
    \end{tabular}
    \caption{Example SQL queries for easy, plug-n-play exploration of the metadata.}
    \label{tab:example_SQL}
\end{table*}

\pagebreak
\section{DB Browser Examples}\label{appendix:browser}
In the section, we provide screenshots of different ways to use DB Browser (\url{https://sqlitebrowser.org/}). Namely, to execute SQL and to filter columns by a single keyword.

\begin{figure*}[ht!]
    \centering
    \includegraphics[width=13cm]{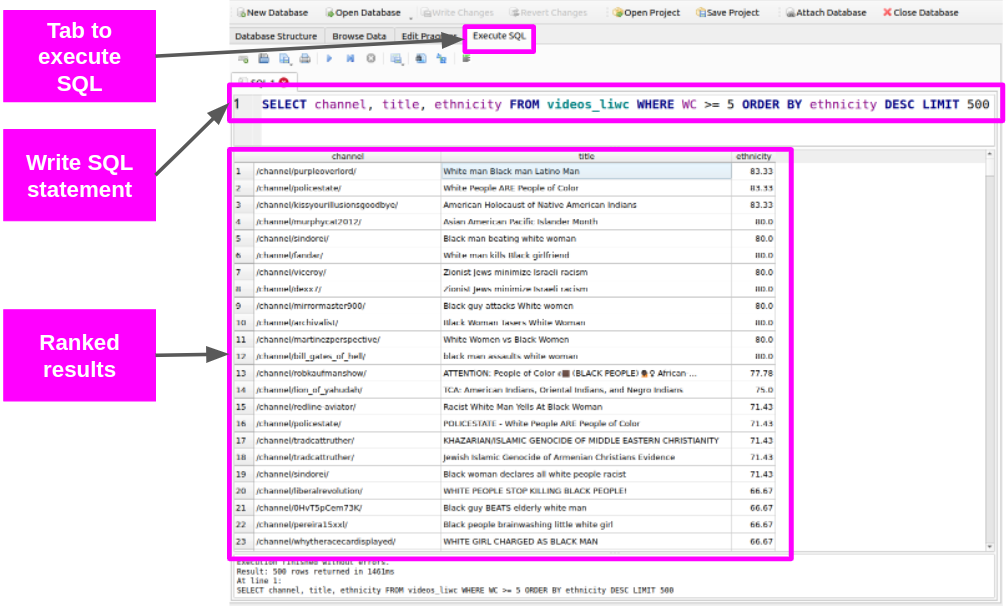}
    \caption{Screenshot of DB Browser SQL query screen. To explore dataset, open database in DB Browser, navigate to the Execute SQL tab, write or copy SQL query into middle box, and press the green play button. One can examine the columns and structure of each table in the database by using the Database Structure tab.}
    \label{fig:dbbrow}
\end{figure*}

\begin{figure*}[ht!]
    \centering
    \includegraphics[width=13cm]{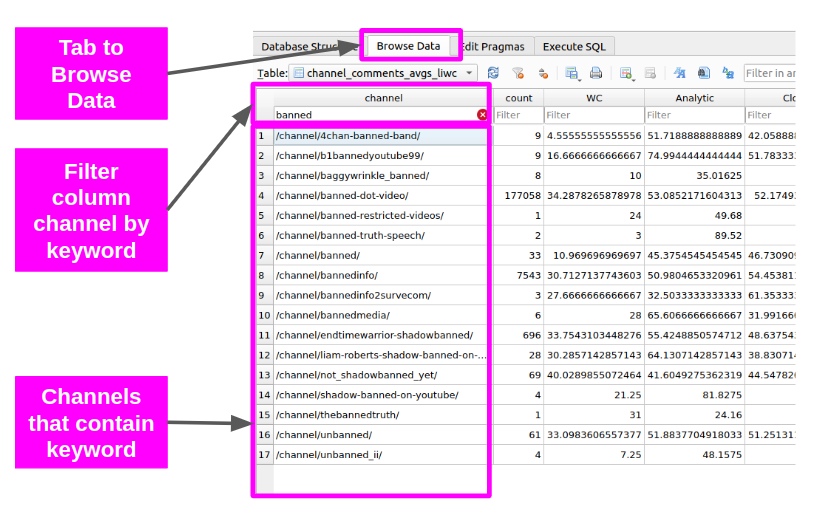}
    \caption{Screenshot of DB Browser `Browse Data'. To explore dataset, open database in DB Browser, navigate to the Browse Data SQL tab, type keyword to filter data by in the desired column. For example, to get all tables with the word `banned' we can filter the channel column.}
    \label{fig:dbbrow}
\end{figure*}

\pagebreak
\section{Example Channel Rankings}\label{appendix:ranking}

\begin{figure*}[ht!]
    \centering
    \begin{subfigure}{.45\textwidth}
        \centering
        \includegraphics[width=7.3cm]{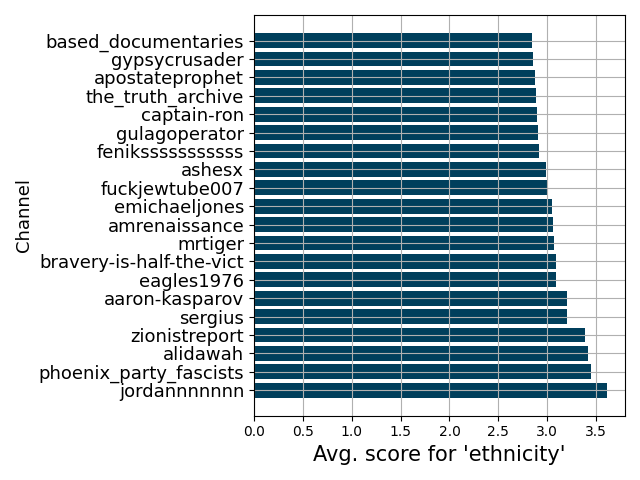}
        \caption{Channels ranked by use of `ethnicity' words in comments}
        \label{fig:ethnicity}
    \end{subfigure}
    \begin{subfigure}{.45\textwidth}
        \centering
        \includegraphics[width=7.3cm]{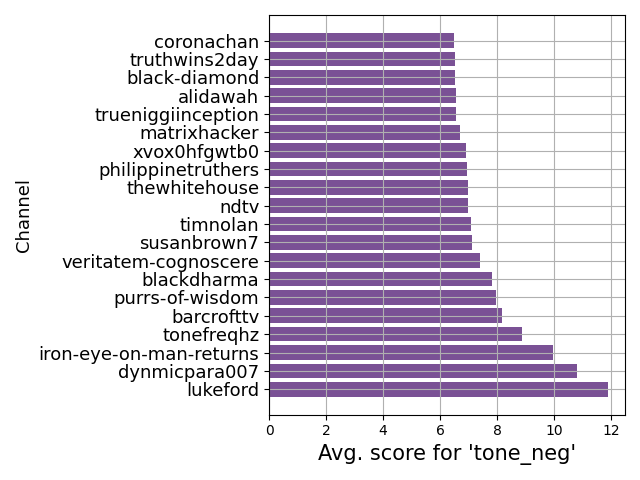}
        \caption{Channels ranked by use of `tone\_neg' words in comments}
        \label{fig:toneneg}
    \end{subfigure}\\
    \begin{subfigure}{.45\textwidth}
        \centering
        \includegraphics[width=7.3cm]{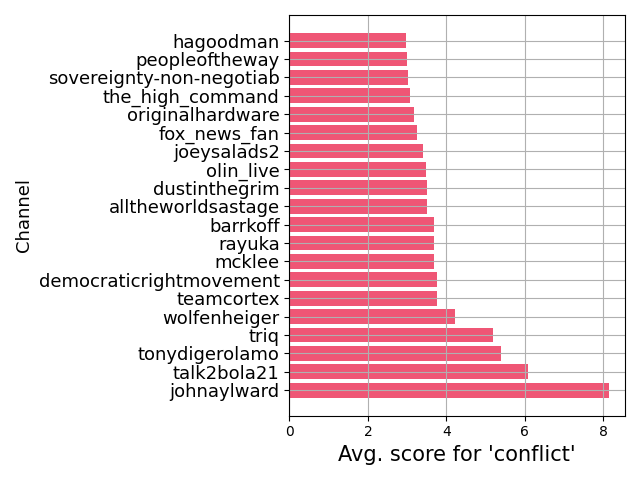}
        \caption{Channels ranked by use of `conflict' words in video titles}
        \label{fig:conflict}
    \end{subfigure}
    \begin{subfigure}{.45\textwidth}
        \centering
        \includegraphics[width=7.3cm]{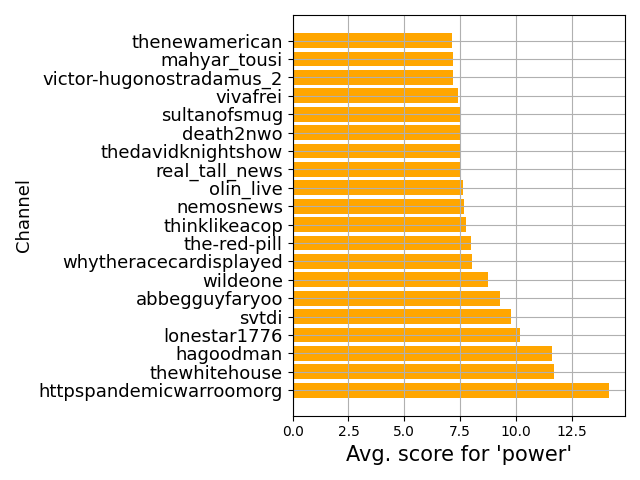}
        \caption{Channels ranked by use of `power' words in video titles}
        \label{fig:power}
    \end{subfigure}
    \begin{subfigure}{.45\textwidth}
        \centering
        \includegraphics[width=7.3cm]{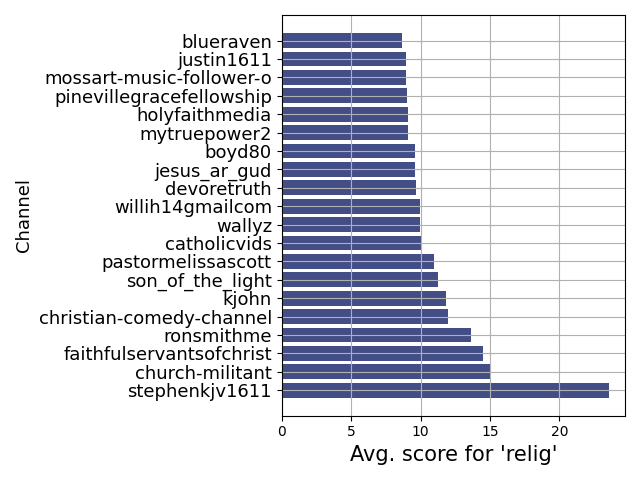}
        \caption{Channels ranked by use of `relig' words in video titles}
        \label{fig:relig}
    \end{subfigure}
    \begin{subfigure}{.45\textwidth}
        \centering
        \includegraphics[width=7.3cm]{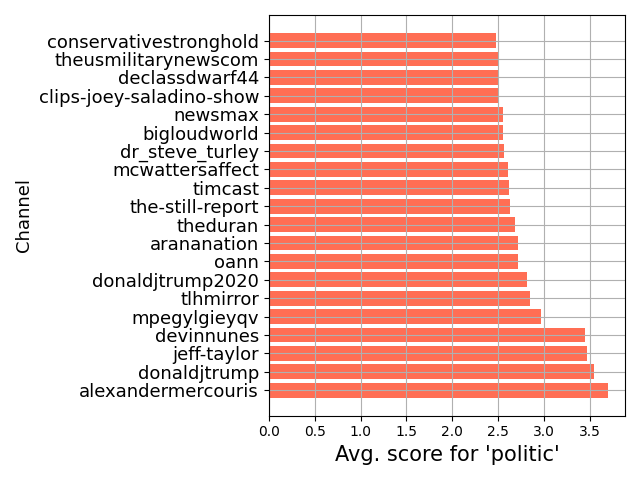}
        \caption{Channels ranked by use of `politic' words in comments}
        \label{fig:politics}
    \end{subfigure}
\caption{In (a) we show the top 20 channels ranked by the use of ethnicity words in comments on average, for channels with at least 500 comments. In (b) we show the top 20 channels ranked by the use of negative tone words in comments on average, for channels with at least 500 comments. In (c) we show the top 20 channels ranked by the use of conflict words in video titles on average, for channels with at least 500 videos. In (d) we show the top 20 channels ranked by the use of power words in video titles on average, for channels with at least 500 videos. In (e) we show the top 20 channels ranked by the use of religion words in video titles on average, for channels with at least 500 videos. In (f) we show the top 20 channels ranked by the use of political/politics words in comments on average, for channels with at least 500 comments. Note, in each we only show the first 25 characters of channel names for the visualization.}
\end{figure*}

\twocolumn
\begin{small}
\bibliography{scibib}

\begin{thebibliography}{29}
\providecommand{\natexlab}[1]{#1}

\bibitem[{Allport(1942)}]{allport1942use}
Allport, G.~W. 1942.
\newblock The use of personal documents in psychological science.
\newblock \emph{Social Science Research Council Bulletin}.

\bibitem[{Althoff, Jindal, and Leskovec(2017)}]{althoff2017online}
Althoff, T.; Jindal, P.; and Leskovec, J. 2017.
\newblock Online actions with offline impact: How online social networks
  influence online and offline user behavior.
\newblock In \emph{Proceedings of the tenth ACM international conference on web
  search and data mining}, 537--546.

\bibitem[{Bantum and Owen(2009)}]{bantum2009evaluating}
Bantum, E.~O.; and Owen, J.~E. 2009.
\newblock Evaluating the validity of computerized content analysis programs for
  identification of emotional expression in cancer narratives.
\newblock \emph{Psychological assessment}, 21(1): 79.

\bibitem[{Boyd et~al.(2022)Boyd, Ashokkumar, Seraj, and
  Pennebaker}]{boyddevelopment}
Boyd, R.~L.; Ashokkumar, A.; Seraj, S.; and Pennebaker, J.~W. 2022.
\newblock The Development and Psychometric Properties of LIWC-22.

\bibitem[{Cannava et~al.(2018)Cannava, High, Jones, and
  Bodie}]{cannava2018stuff}
Cannava, K.~E.; High, A.~C.; Jones, S.~M.; and Bodie, G.~D. 2018.
\newblock The stuff that verbal person-centered support is made of: Identifying
  linguistic markers of more and less supportive conversations.
\newblock \emph{Journal of Language and Social Psychology}, 37(6): 656--679.

\bibitem[{Coppersmith, Harman, and Dredze(2014)}]{coppersmith2014measuring}
Coppersmith, G.; Harman, C.; and Dredze, M. 2014.
\newblock Measuring post traumatic stress disorder in Twitter.
\newblock In \emph{Eighth international AAAI conference on weblogs and social
  media}.

\bibitem[{Crawford et~al.(2015)Crawford, Khoshgoftaar, Prusa, Richter, and
  Al~Najada}]{crawford2015survey}
Crawford, M.; Khoshgoftaar, T.~M.; Prusa, J.~D.; Richter, A.~N.; and Al~Najada,
  H. 2015.
\newblock Survey of review spam detection using machine learning techniques.
\newblock \emph{Journal of Big Data}, 2(1): 1--24.

\bibitem[{del Pilar Salas-Z{\'a}rate et~al.(2014)del Pilar Salas-Z{\'a}rate,
  L{\'o}pez-L{\'o}pez, Valencia-Garc{\'\i}a, Aussenac-Gilles, Almela, and
  Alor-Hern{\'a}ndez}]{del2014study}
del Pilar Salas-Z{\'a}rate, M.; L{\'o}pez-L{\'o}pez, E.; Valencia-Garc{\'\i}a,
  R.; Aussenac-Gilles, N.; Almela, {\'A}.; and Alor-Hern{\'a}ndez, G. 2014.
\newblock A study on LIWC categories for opinion mining in Spanish reviews.
\newblock \emph{Journal of Information Science}, 40(6): 749--760.

\bibitem[{Donovan, Lewis, and Friedberg(2019)}]{donovan2019parallel}
Donovan, J.; Lewis, B.; and Friedberg, B. 2019.
\newblock Parallel ports: Sociotechnical change from the alt-right to alt-tech.

\bibitem[{Eichstaedt et~al.(2018)Eichstaedt, Smith, Merchant, Ungar, Crutchley,
  Preo{\c{t}}iuc-Pietro, Asch, and Schwartz}]{eichstaedt2018facebook}
Eichstaedt, J.~C.; Smith, R.~J.; Merchant, R.~M.; Ungar, L.~H.; Crutchley, P.;
  Preo{\c{t}}iuc-Pietro, D.; Asch, D.~A.; and Schwartz, H.~A. 2018.
\newblock Facebook language predicts depression in medical records.
\newblock \emph{Proceedings of the National Academy of Sciences}, 115(44):
  11203--11208.

\bibitem[{Flaxman, Goel, and Rao(2016)}]{flaxman2016filter}
Flaxman, S.; Goel, S.; and Rao, J.~M. 2016.
\newblock Filter bubbles, echo chambers, and online news consumption.
\newblock \emph{Public opinion quarterly}, 80(S1): 298--320.

\bibitem[{Francis and Booth(1993)}]{francis1993linguistic}
Francis, M.; and Booth, R.~J. 1993.
\newblock Linguistic inquiry and word count.
\newblock \emph{Southern Methodist University: Dallas, TX, USA}.

\bibitem[{Guess et~al.(2018)Guess, Nyhan, Lyons, and
  Reifler}]{guess2018avoiding}
Guess, A.; Nyhan, B.; Lyons, B.; and Reifler, J. 2018.
\newblock Avoiding the echo chamber about echo chambers.
\newblock \emph{Knight Foundation}, 2: 1--25.

\bibitem[{Hirsh and Peterson(2009)}]{hirsh2009personality}
Hirsh, J.~B.; and Peterson, J.~B. 2009.
\newblock Personality and language use in self-narratives.
\newblock \emph{Journal of research in personality}, 43(3): 524--527.

\bibitem[{Horne and Adali(2017)}]{horne2017just}
Horne, B.; and Adali, S. 2017.
\newblock This just in: Fake news packs a lot in title, uses simpler,
  repetitive content in text body, more similar to satire than real news.
\newblock In \emph{Proceedings of the international AAAI conference on web and
  social media}, volume~11, 759--766.

\bibitem[{Jasser et~al.(2021)Jasser, McSwiney, Pertwee, and
  Zannettou}]{jasser2021welcome}
Jasser, G.; McSwiney, J.; Pertwee, E.; and Zannettou, S. 2021.
\newblock ‘Welcome to\# GabFam’: Far-right virtual community on Gab.
\newblock \emph{New Media \& Society}, 14614448211024546.

\bibitem[{Larcker and Zakolyukina(2012)}]{larcker2012detecting}
Larcker, D.~F.; and Zakolyukina, A.~A. 2012.
\newblock Detecting deceptive discussions in conference calls.
\newblock \emph{Journal of Accounting Research}, 50(2): 495--540.

\bibitem[{Munn(2021)}]{munn2021more}
Munn, L. 2021.
\newblock More than a mob: Parler as preparatory media for the US Capitol
  storming.
\newblock \emph{First Monday}.

\bibitem[{Pennebaker, Booth, and Francis(2007)}]{pennebaker2007linguistic}
Pennebaker, J.; Booth, R.; and Francis, M. 2007.
\newblock Linguistic Inquiry and Word Count: LIWC2007—Operator’s Manual.
  LIWC. net.

\bibitem[{Pennebaker et~al.(2015)Pennebaker, Boyd, Jordan, and
  Blackburn}]{pennebaker2015development}
Pennebaker, J.~W.; Boyd, R.~L.; Jordan, K.; and Blackburn, K. 2015.
\newblock The development and psychometric properties of LIWC2015.
\newblock Technical report.

\bibitem[{Pennebaker et~al.(2014)Pennebaker, Chung, Frazee, Lavergne, and
  Beaver}]{pennebaker2014small}
Pennebaker, J.~W.; Chung, C.~K.; Frazee, J.; Lavergne, G.~M.; and Beaver, D.~I.
  2014.
\newblock When small words foretell academic success: The case of college
  admissions essays.
\newblock \emph{PloS one}, 9(12): e115844.

\bibitem[{Pennebaker, Francis, and Booth(2001)}]{pennebaker2001linguistic}
Pennebaker, J.~W.; Francis, M.~E.; and Booth, R.~J. 2001.
\newblock Linguistic inquiry and word count: LIWC 2001.
\newblock \emph{Mahway: Lawrence Erlbaum Associates}, 71(2001): 2001.

\bibitem[{Rice et~al.(2022)Rice, Horne, Luther, Borycz, Allard, Ruck,
  Fitzgerald, Manaev, Prins, Taylor et~al.}]{rice2022monitoring}
Rice, N.~M.; Horne, B.~D.; Luther, C.~A.; Borycz, J.~D.; Allard, S.~L.; Ruck,
  D.~J.; Fitzgerald, M.; Manaev, O.; Prins, B.~C.; Taylor, M.; et~al. 2022.
\newblock Monitoring event-driven dynamics on Twitter: a case study in Belarus.
\newblock \emph{SN Social Sciences}, 2(4): 1--20.

\bibitem[{Schwartz et~al.(2013)Schwartz, Eichstaedt, Kern, Dziurzynski, Lucas,
  Agrawal, Park, Lakshmikanth, Jha, Seligman
  et~al.}]{schwartz2013characterizing}
Schwartz, H.; Eichstaedt, J.; Kern, M.; Dziurzynski, L.; Lucas, R.; Agrawal,
  M.; Park, G.; Lakshmikanth, S.; Jha, S.; Seligman, M.; et~al. 2013.
\newblock Characterizing geographic variation in well-being using tweets.
\newblock In \emph{Proceedings of the International AAAI Conference on Web and
  Social Media}, volume~7, 583--591.

\bibitem[{Shu, Wang, and Liu(2019)}]{shu2019beyond}
Shu, K.; Wang, S.; and Liu, H. 2019.
\newblock Beyond news contents: The role of social context for fake news
  detection.
\newblock In \emph{Proceedings of the twelfth ACM international conference on
  web search and data mining}, 312--320.

\bibitem[{Trujillo et~al.(2020)Trujillo, Gruppi, Buntain, and
  Horne}]{trujillo2020bitchute}
Trujillo, M.; Gruppi, M.; Buntain, C.; and Horne, B.~D. 2020.
\newblock {What is BitChute? Characterizing the "Free Speech" Alternative to
  YouTube}.
\newblock In \emph{Proceedings of the 31st ACM Conference on Hypertext and
  Social Media}, HT '20, 139--140. New York, NY, USA: Association for Computing
  Machinery.
\newblock ISBN 9781450370981.

\bibitem[{Trujillo et~al.(2022)Trujillo, Gruppi, Buntain, and
  Horne}]{trujillo2022mela}
Trujillo, M.; Gruppi, M.; Buntain, C.; and Horne, B.~D. 2022.
\newblock The MeLa BitChute Dataset.
\newblock \emph{Proceedings of the International AAAI Conference on Web and
  Social Media}, 16.

\bibitem[{Wilson and Starbird(2021)}]{Wilson2021}
Wilson, T.; and Starbird, K. 2021.
\newblock {Cross-platform Information Operations: Mobilizing Narratives and
  Building Resilience through both 'Big' and 'Alt' Tech}.
\newblock \emph{Proceedings of the ACM on Human-Computer Interaction},
  5(CSCW2).

\bibitem[{Zannettou et~al.(2020)Zannettou, Finkelstein, Bradlyn, and
  Blackburn}]{zannettou2020quantitative}
Zannettou, S.; Finkelstein, J.; Bradlyn, B.; and Blackburn, J. 2020.
\newblock A quantitative approach to understanding online antisemitism.
\newblock In \emph{Proceedings of the International AAAI Conference on Web and
  Social Media}, volume~14, 786--797.

\end{thebibliography}
\end{small}

\end{document}